\documentclass[letterpaper, 10 pt, conference]{ieeeconf}  
\usepackage{graphicx}
\IEEEoverridecommandlockouts                              
\usepackage{amsmath}
\usepackage{amssymb}
\usepackage{bm}
\usepackage{graphics}
\usepackage[hidelinks]{hyperref}
\usepackage{xcolor}
\usepackage{dblfloatfix}
\usepackage {tikz}
\usepackage{mathtools}
\usepackage{svg}
\usepackage[bottom]{footmisc}
\usepackage{amsmath}
\usepackage{soul}
\usepackage[switch]{lineno}
\usepackage{multirow}
\usepackage{caption}
\usepackage{tabularray}
\usepackage[noend]{algpseudocode}
\usepackage[linesnumbered,ruled,vlined, noend]{algorithm2e}
\usetikzlibrary {positioning}

\title{\Large \bf Collaborative Adaptation for Recovery from Unforeseen Malfunctions \\in Discrete and Continuous MARL Domains
}

\author{Yasin Findik, Hunter Hasenfus, and Reza Azadeh
\thanks{Authors are with the Persistent Autonomy and Robot Learning (PeARL) Lab, Richard Miner School of  Computer and Information Sciences, University of Massachusetts Lowell, MA, USA {\tt\small yasin\_findik@student.uml.edu, reza@cs.uml.edu}}}

\begin{document}
\maketitle
\thispagestyle{empty}
\pagestyle{empty}

\begin{abstract}

Cooperative multi-agent learning plays a crucial role for developing effective strategies to achieve individual or shared objectives in multi-agent teams. In real-world settings, agents may face unexpected failures, such as a robot's leg malfunctioning or a teammate's battery running out. These malfunctions decrease the team's ability to accomplish assigned task(s), especially if they occur after the learning algorithms have already converged onto a collaborative strategy. Current leading approaches in Multi-Agent Reinforcement Learning (MARL) often recover slowly -- if at all -- from such malfunctions. To overcome this limitation, we present the Collaborative Adaptation (CA) framework, highlighting its unique capability to operate in both continuous and discrete domains. Our framework enhances the adaptability of agents to unexpected failures by integrating inter-agent relationships into their learning processes, thereby accelerating the recovery from malfunctions. We evaluated our framework's performance through experiments in both discrete and continuous environments. Empirical results reveal that in scenarios involving unforeseen malfunction, although state-of-the-art algorithms \textit{often} converge on sub-optimal solutions, the proposed CA framework mitigates and recovers more \textit{effectively}.

\end{abstract}

\section{Introduction}

The popularity of multi-agent learning methods is rising due to their ability to manage the complexities of robotic systems and their importance in tasks that require detailed coordination~\cite{dorri2018multi}. These methods are valuable across various fields, such as robotics for search \& rescue operations~\cite{kleiner2006rfid} and autonomous driving~\cite{pendleton2017perception}, whether involving multiple or single robot systems. In single-robot applications, multi-agent learning facilitates coordination among different system modules, exemplified by treating each joint of a robot manipulator as an individual agent within a collaborative team. In both multi-robot and single-robot systems, the collaborative behavior of the agents is particularly crucial for quickly and autonomously recovering from unexpected malfunctions, such as joint failures or battery failure that prevents the robot(s) from moving. In these situations, agents must improve their collaboration further and modify their joint strategies to successfully tackle these challenges.

In the field of Multi-Agent Reinforcement Learning (MARL) for cooperative tasks, the Centralized Training with Decentralized Execution (CTDE)~\cite{oliehoek2008optimal} paradigm has emerged as a prominent approach, notably through recent advancements in deep reinforcement learning~\cite{huttenrauch2019deep, levine2016end, mnih2015human}. It effectively addresses a range of cooperative challenges in MARL, including the curse of dimensionality~\cite{shoham2007if}, non-stationarity~\cite{busoniu2008comprehensive} and global exploration~\cite{matignon2012independent}. However, despite its effectiveness in collaborative tasks, CTDE methods struggle with rapid adaptation to unforeseen agent malfunction(s) or failure. This challenge might be due to two primary reasons: the lack of built-in mechanisms to manage unforeseen robot failures, or the absence of design elements to steer agents' collaboration strategies. Consequently, these algorithms face delays in adapting, as they need to independently discover and converge on new \textit{effective} collaboration strategies without any guidance. To assist agents in searching for these strategies, we associate the discovery of such collaborative behaviors through inter-agent relationships, thereby accelerating and improving the adaptability of the agents as a team.

In this paper, we introduce a novel MARL framework that integrates a relational network with any multi-agent learning algorithm following the CTDE paradigm. This integration highlights the relative importance among agents, thereby enhancing their collaboration towards a specific strategy. This innovation facilitates faster and more effective adaptation to unexpected robot failures. We conducted experiments in two settings: a multi-agent discrete environment with multiple robots and a continuous environment~\cite{gymnasium_robotics2023github} simulating an ant with multiple agents (an agent per leg). We demonstrate the effectiveness of our approach by comparing it with Independent Deep Q-Networks (IDQN)~\cite{tampuu2017multiagent} and Value Decomposition Networks (VDN)~\cite{sunehag2017value} in the discrete environment, and with Independent Q-Functionals (IQF)~\cite{findik2024mixed} and Multi-Agent Deep Deterministic Policy Gradient (MADDPG)~\cite{lowe2017multi} in the continuous environment. Empirical results indicate that our method improves teamwork among agents, whether in configurations like a single robot with multiple legs/joints or multiple robots. More importantly, it enhances the ability to adapt to unexpected failure by utilizing relational networks.

\section{Background and Related Work}

Reinforcement learning (RL) aims at enabling an agent to maximize its cumulative future rewards through learning from its interactions within an environment. This process is often represented as a  Decentralized Markov Decision Process (DEC-MDP), characterized by a tuple $\langle  \mathcal{S}, \mathcal{A}, \mathcal{R}, \mathcal{T}, \gamma, N\rangle$. $\mathcal{S}$, $\mathcal{A}$ and $\mathcal{R}$ indicate the true state of the environment, the joint set of agents' action and reward, respectively. $\mathcal{T} (s, \mathcal{A}, s') \colon \mathcal{S} \times \mathcal{A}\times \mathcal{S} \mapsto [0,1]$ is the dynamics function, defining the transition probability, where $s$ and $s'$ are the current and next state, respectively. $\gamma\in[0,1)$ is the discount factor, and $N$ represents the number of agents.

\vspace{.2cm}
\textbf{Q-Learning \& Q-Functionals.} Q-Learning~\cite{watkins1992q} and Deep Q-Networks (DQN)~\cite{mnih2015human} are two prominent \textit{value-based} approaches in single-agent RL with discrete action spaces. Q-Learning employs an action-value \textit{table}, $Q^{\pi}$, for a policy $\pi$, defined as $Q^{\pi}(s, a)=\mathbb{E}[G| s^t=s, a^t=a]$, where $G$ and $t$ denote the return and time-step, respectively. This can be re-formulated recursively as:
\begin{align*}
Q^{\pi}(s, a)=\mathbb{E}_{s'}[R(s,a) + \gamma \mathbb{E}_{a'\sim\pi}[Q^{\pi}(s', a')]],
\end{align*}
\noindent 
where $R$ represents the reward function. However, DQN learns an action-value \textit{function}, denoted as $\hat{Q}$, corresponding to the optimal policy by minimizing this loss:
\begin{align}
\label{loss_DQN}
\resizebox{.91\linewidth}{!}{$
            \displaystyle
            L(w_{\textrm{PN}})=\mathbb{E}_{s,a,r,s'}[(r + \gamma \max_{a'}(\hat{Q}(s', a'; w_{\textrm{TN}}))- \hat{Q}(s, a; w_{\textrm{PN}}))^2].
        $}
\end{align}
\noindent 
The parameters for the prediction network are denoted by $w_{\textrm{PN}}$, while $w_{\textrm{TN}}$ represents the parameters for the target network. $w_{\textrm{TN}}$ are periodically updated with $w_{\textrm{PN}}$ to enhance the stability of the learning process. DQN also utilizes an experience replay memory that hold tuples $\langle s, a, r, s'\rangle$, to enhance the learning stability further.

In these approaches, state-action pairs are directly associated with $\mathcal{R}$, as denoted by:
\begin{align}
\label{tqa}
\hat{Q}(s,a): (\mathcal{S} \times \mathcal{A}) \mapsto \mathcal{R}.
\end{align}
\noindent
This direct mapping restricts their utility to scenarios within discrete environments (i.e., a finite set of actions) because it is unfeasible to calculate the action-value for every state-action pair in continuous domains. To address this limitation, Q-Functionals (QF)~\cite{lobel2023q}, restructure the mapping by decoupling the state and action elements, illustrated as: 
\begin{align}
\label{fqa}
\hat{Q}^{\textrm{F}}(s,a): \mathcal{S} \mapsto (\mathcal{A} \mapsto \mathcal{R}).
\end{align}
\noindent
Here, the state is initially mapped to a function that subsequently associates actions with $\mathcal{R}$. Essentially, each state is depicted as a function formed through the learning of basis function coefficients within the action domain. These state-representative functions enable the quick assessment of many actions via matrix operations between the action representations and the learned coefficients. Thus, QF adeptly manage continuous domains in single-agent contexts.

Within multi-agent systems, the basic application of these value-based approaches involves each agent $i$ independently optimizing its respective action-value table or function, such as Independent Q-Learning (IQL)~\cite{tan1993multi}, Independent Deep Q-Networks (IDQN)~\cite{mnih2015human}, Independent Q-Functionals (IQF)~\cite{findik2024mixed}. These methods, however, encounter difficulties because the independent policy modifications by each agent make the environment appear non-stationary from their respective perspectives. This non-stationarity violates the stationary requirements of the Markov property, crucial for the stable convergence of these algorithms. As a simple alternative to independent learning, fully centralized learning~\cite{claus1998dynamics} can be adopted, in which a singular controller is utilized across all agents, enabling the joint learning of a value function. Yet, this centralized strategy is characterized by computational demand and scalability problems, as the observation and action spaces grow exponentially with the number of agents, possibly resulting in intractability.

\vspace{.2cm}
\textbf{Value Function Factorization.} 
In the context of joint action-value function learning, where complexity escalates exponentially with the number of agents, value function factorization methods emerge as a promising solution. Following the centralized training with decentralized learning (CTDE)~\cite{oliehoek2008optimal} paradigm, these methods facilitate autonomous action execution by individual agents, while centralizing the integration of their strategies. Consequently, these methods resolve the non-stationarity issue inherent in independent learning through centralized training and tackle the scalability challenges associated with centralized learning by enabling decentralized execution.

VDN~\cite{sunehag2017value} and QMIX~\cite{rashid2020monotonic} stand out as two significant learning algorithms in action-value function factorization. These algorithms are characterized by using separate action-value function, denoted as \smash{$\hat{Q}_i$}, for each agent $i \in \{ 1,...,N\}$. The primary distinction between these methods is their strategy for calculating the total action value, $Q_{\textrm{tot}}$. VDN achieves $Q_{\textrm{tot}}$ by summing the $Q_i$s, as expressed in: 
\begin{align}
\label{VDN}
\hat{Q}_{\textrm{tot}} = \sum_{i=1}^{N} \hat{Q}_i(s, a_i),
\end{align}
whereas QMIX employs a state-dependent continuous monotonic function to integrate these values, denoted by:
\begin{align}
\label{QMIX}
\hat{Q}_{\textrm{tot}} \coloneqq f_s(\hat{Q}_1(s, a_1), ..., \hat{Q}_n(s, a_N)),  
\end{align}
with the condition of $\frac{\partial f_s}{\partial \hat{Q}_i} \ge 0,  \forall i \in \{1, ..., N\}$. Often, these factorization algorithms use DQN for approximating the action-value function and aim to minimize the loss~\eqref{loss_DQN} in a centralized way, as follows:
\begin{align}
\label{td_error_vdn}
\resizebox{.91\linewidth}{!}{$
            \displaystyle
            L(w_{\textrm{PN}})=\mathbb{E}_{s,u,r_{\textrm{team}},s'}[(r_{\textrm{team}} + \gamma \max_{u'}(\hat{Q}_{\textrm{tot}}(s', u'; w_{\textrm{TN}}))- \hat{Q}_{\textrm{tot}}(s, u; w_{\textrm{PN}}))^2].
        $}
\end{align}
\noindent
where $r_{\textrm{team}}$ is obtained by uniformly summing the agents individual rewards, and $u$ is the joint action of the agents.

The current value function factorization methods have demonstrated proficiency in creating stable collaborative within discrete environments. However, even with their advanced iterations~\cite{findik2023impact, findik2023influence}, these approaches encounter challenges in formulating effective policies within continuous environments.  In such settings, the action space is delineated by continuous vectors, thereby encompassing an unbounded set of potential actions~\cite{lim2018actor}.

\vspace{.2cm}
\textbf{Policy Gradient (PG).}
In single-agent RL with continuous action spaces Policy Gradient (PG) approaches~\cite{sutton1999policy} are essential. These methods aim to directly optimize the policy parameters, symbolized as $\theta$, to maximize the expected return. The principal strategy is to modify $\theta$ in the direction of the policy's gradient, formulated as:
\begin{align}
\label{PG}
\nabla_\theta J (\theta)= \mathbb{E}_{s \sim p^{\pi}, a \sim \pi_{\theta}}[\nabla_\theta\log \pi_{\theta} (a|s)Q^\pi(s,a)],
\end{align}
\noindent
where $p^{\pi}$ denotes the distribution of states under the policy $\pi$. 
Generally, PG approaches differ from each other in their methods for estimating $Q^\pi$.

Moreover, Deterministic Policy Gradient (DPG)~\cite{silver2014deterministic} adapts the PG theorem and represents policy as $\mu_{\theta}: \mathcal{S} \mapsto \mathcal{A}$ with a vector of $n$ parameters $\theta \in \mathbb{R}^n$. The formulation of policy gradient becomes as:
\begin{align}
\label{DPG}
\nabla_\theta J (\mu_\theta)= \mathbb{E}_{s \sim p^{\mu}}[\nabla_\theta\mu_{\theta}(s) \nabla_{a}Q^{\mu}(s,a)|_{a=\mu_{\theta}(s)}].   
\end{align}
\noindent
Also, Deep Deterministic Policy Gradient (DDPG)~\cite{lillicrap2015continuous} approach extends DPG by employing deep neural networks to approximate both the policy $\mu$ and critic \smash{$\hat{Q}^\mu$}. 

\vspace{.2cm}
\textbf{Multi-Agent DDPG (MADDPG).} 
MADDPG~\cite{lowe2017multi} extends DDPG by adhering CTDE paradigm, facilitating its application in multi-agent settings. The policy of agents are represented as $\bm{\pi}=[\pi_1, \dots, \pi_N]$ with parameters $\bm{\theta} = [\theta_1, \dots, \theta_n]$ in MADDPG. The optimization of these policies is conducted through the gradient of the expected return, which for each agent $i \in \{ 1,...,N\}$ is formulated as:
\begin{align*}
\resizebox{.95\linewidth}{!}{$
            \displaystyle
            \nabla_{\theta_i} J (\theta_i)= \mathbb{E}_{x \sim p^{\mu}, a_i \sim \pi_{i}}[\nabla_{\theta_i} \log \pi_{i} (a_i|o_i)\hat{Q}_i^{\bm{\pi}}(\bm{o}, \bm{a})],
        $}
\end{align*}
\noindent
where \smash{$\hat{Q}_i^\pi(\bm{o}, \bm{a})$} denotes the centralized action-value function, incorporating both the states $\bm{o}=[o_1,\dots,o_N]$ and actions $\bm{a}=[a_1,\dots,a_N]$ of all agents to compute the $Q$-value. The framework is further adapted for deterministic policies $\bm{\mu}=[\mu_{\theta_1}, \dots, \mu_{\theta_N}]$, modifying the gradient expression as:
\begin{align*}
\resizebox{.95\linewidth}{!}{$
            \displaystyle
            \nabla_{\theta_i} J (\mu_{\theta_i}) = \mathbb{E}_{\bm{o} \sim p^{\bm{\mu}}, \bm{a}\sim \bm{\mu}}[\nabla_{\theta_i}\mu_{\theta_i}(a_i|o_i) \nabla_{a_i}\hat{Q}_i^{\bm{\mu}}(\bm{o}, \bm{a})|_{a_i=\mu_{\theta_i}(o_i)}].
        $}
\end{align*}
MADDPG, similar to the DQN algorithm, incorporates target networks and experience replay to enhance the stability of the policy networks. The loss function for updating the centralized action-value function $Q_i^{\bm{\mu}}$ is defined as:
\begin{align*}
\resizebox{.98\linewidth}{!}{$
            \displaystyle
            L(\theta_{i})=\mathbb{E}_{\bm{o},\bm{a},\bm{r},\bm{o'} \sim \mathcal{D}}[(r_i + \gamma Q_i^{\bm\mu'}(\bm{o'},\bm{a'} )|_{\bm{a'}=\bm\mu'(\bm{o'})} - Q_i^{\bm\mu}(\bm{o},\bm{a} ))^2], 
        $}
\end{align*}
\noindent
where $\mathcal{D}$ denotes the replay memory storing tuples $\langle\bm{o}, \bm{a}, \bm{r}, \bm{o'}\rangle$ and $\bm{\mu'} = [\mu_{\theta'_1}, \dots, \mu_{\theta'_N}]$ represents the periodically updated target policies. Despite MADDPG's state-of-the-art performance in maximizing agents' return, it faces challenges like converging to local optima due to inefficient sampling and limited exploration in complex environments with extensive state and action spaces, specifically recovering from unforeseen malfunction(s) or failure.

\section{Proposed Method}

In cooperative MARL, both value-based and policy-based techniques are utilized to maximize collective rewards, aiming to converge toward an optimal solution, potentially the global optimum. The convergence process is influenced by the stochastic nature of agents’ exploration, particularly in scenarios where multiple cooperation strategies offer the same maximum total reward. Thus, the team's collaboration strategy, and inherently its structural dynamics, are significantly influenced by the randomness in agents' exploration process. However, in real-world applications, the importance of the team's structure becomes pronounced when robots encounter mechanical issues (e.g., battery or joint failures), necessitating a strategy that does not solely depend on randomness. In other words, the current state-of-the-art methods demonstrate limited effectiveness in adapting team behavior to scenarios that involve unforeseen failures.

To effectively overcome these challenges, we propose a novel framework known as Collaborative Adaptation (CA). This framework is tailored to consider the dynamics between agents, guiding them toward adopting cooperative strategies to improve team performance and accelerate adaptation. In the current field of cooperative MARL, such a mechanism is notably absent, leading to increased difficulty and time required for adaptation in the face of unforeseen malfunctions. Our CA framework aims to fill this gap by encouraging agents, through inter-agent relationships, to either assist malfunctioning one directly or completing its task on behalf of it, thus improving the team's overall performance.

\renewcommand{\algorithmicrequire}{\textbf{Input:}}
\renewcommand{\algorithmicensure}{\textbf{Output:}}

\begin{algorithm}[t]
    \caption{Collaborative Adaptation}
    \label{alg:CA}
    \SetAlgoLined
    \SetKwInOut{Input}{input}
    \SetKwInOut{Output}{output}
    \DontPrintSemicolon
    
    \Input{P-NN, $\hat{Q}^{\textrm{prediction}}$; T-NN, $\hat{Q}^{\textrm{target}}$; relational network, $G$; batch size, $b$; number of iterations for updates, $m$
    }
    \ForEach{episode}{
        Initialize $s$\;
        \ForEach{step of episode}{
            Choose $a$ from $s$ using policy derived from $\hat{Q}^{\textrm{prediction}}$ (with $\varepsilon$-greedy)\;
            Take action $a$, observe $r$, $s'$\;
            Store $s$, $a$, $r$, $s'$ in memory\;
            $s \leftarrow s'$\;
        }
        \For{$i=1, \ldots, m$}{
            $S$, $A$, $R$, $S'$ $\leftarrow$ sample chunk, size of $b$, from memory\;
            $Q^\textrm{prediction}_\textrm{values} \leftarrow \hat{Q}^{\textrm{prediction}}$($S$) \;
            $Q^\textrm{prediction}_\textrm{values} \leftarrow$ action $A$ of $Q^\textrm{prediction}_\textrm{values}$ of every agent in every sample\;
            $Q^{\textrm{prediction}}_{\textrm{total}} \leftarrow$ use~\eqref{q_team_G} with $G$ and $Q^{\textrm{prediction}}$\;
            $Q^\textrm{target}_\textrm{values} \leftarrow \hat{Q}^{\textrm{target}}$($S'$)\;
            $Q^\textrm{target} \leftarrow$ max of $Q^\textrm{target}_\textrm{values}$ of every agent in every sample \;
            $Q^{\textrm{target}}_{\textrm{total}} \leftarrow$ use~\eqref{q_team_G} with $G$ and $Q^{\textrm{target}}$\;
            $R^{\textrm{team}} \leftarrow$ use~\eqref{r_team_G} with $G$ and $R$\;

            $L \leftarrow$ use~\eqref{td_error_vdn} with $R^{\textrm{team}}$, $Q^{\textrm{target}}_{\textrm{total}}$, $Q^{\textrm{prediction}}_{\textrm{total}}$\;
        
            Backpropagate $L$ to the weights of $\hat{Q}^{\textrm{prediction}}$\;
        }
        Update the parameters of $\hat{Q}^{\textrm{target}}$ with the parameters of $\hat{Q}^{\textrm{prediction}}$ 
    }
\end{algorithm}

The proposed framework integrates a relational network into the agents' learning process, structured as a directed graph $\mathcal{G}=(\mathcal{V}, \mathcal{E}, \mathcal{W})$, which defines the inter-agent relationships to enhance cooperative strategies. 
\begin{center}
\begin {tikzpicture}[-latex,auto ,node distance =3 cm and 2.5cm ,on grid , semithick, state/.style ={ circle, top color=white ,  draw, black, bottom color = black!20, text=black, minimum width =0.5 cm}]
\node[state] (A) []  {$v_1$};
\node[state] (B) [right =of A] {$v_2$};
\path (A) edge [loop left] node[left] {$w_{11}=0.3$} (A);
\path (B) edge [loop right] node[right] {$w_{22}=1$} (B);
\path (A) edge node[above =0.01 cm] {$w_{12}=0.7$} (B);
\end{tikzpicture}
\end{center}
\noindent
Each agent $i \in \{1, 2, ..., N\}$ is depicted as a vertex $v_i$ in $\mathcal{V}$, with $\mathcal{E}$ representing the set of directed edges $e_{ij}$ extending from $v_i$ to $v_j$, and $\mathcal{W}$ containing the edge weights $w_{ij}\in[0, 1]$. These weights, $w_{ij}$, quantify the level of influence or interest agent $i$ has in the outcomes of agent $j$.

In the context of the assigned tasks to agents, the relationships may be utilized either for the computation of the team's action value, $Q_{\textrm{tot}}$, through the agents' individual action-values, $Q_i$, or for the calculation of the team reward, $r_{\textrm{team}}$, via the individual rewards of the agents, $r_i$. For instance, in scenarios where agents receive rewards based on individual actions while engaging in a collective task, the aggregate team reward is calculated for use in \eqref{td_error_vdn}, as follows:
\begin{align}
r_{\textrm{team}} = \sum_{i\in\mathcal{V}} \sum_{j\in\mathcal{E}i} w_{ij}r_j.
\label{r_team_G}
\end{align}
\noindent
In cases where agents receive a uniform team reward from the environment and lack access to individual rewards, the team's action-value, $Q_{\textrm{tot}}$, is re-defined to incorporate the inter-agent relationships for utilization in \eqref{td_error_vdn}, as:  
\begin{align}
Q_{\textrm{tot}} = \sum_{i\in\mathcal{V}}^{} \sum_{j\in\mathcal{E}_i}^{} w_{ij}Q_j,
\label{q_team_G}
\end{align}
\noindent
where $Q_j$ denotes the action-value of agent $j$.

The modifications introduced in our framework allow for the direction or guidance of the agents' collaborative approach, thereby improving their adaptability to unexpected robot failure and accelerating the convergence to optimal solutions even without any malfunction, as evidenced by our empirical findings. During our experimental investigations, VDN~\cite{rashid2020monotonic} and Mixed Q-Functionals (MQF)~\cite{findik2024mixed} were employed as the learning algorithms for the agents within the CA framework, tailored for discrete and continuous settings, denoted as CA-VDN and CA-MQF, respectively. The pseudo-code corresponding to the CA framework is presented in Algorithm~\ref{alg:CA}.

\section{Environments}

To evaluate how effectively our framework influences agent behaviors and improves their ability to adapt to unexpected failures of their teammate(s) in both discrete and continuous tasks, we conducted experiments in two distinct environments. In the context of discrete tasks, we employed a multi-agent grid-world environment and compared our framework's performance against value-based methods, particularly IDQN and VDN. For continuous tasks, we used a multi-agent MuJoCo environment and conducted comparisons with policy-based methods, specifically variants of DDPG designed for multi-agent scenarios. 

\subsection{Multi-agent Grid-world Environment}

\begin{figure}[t]
\centering
  \includegraphics[width=\linewidth]{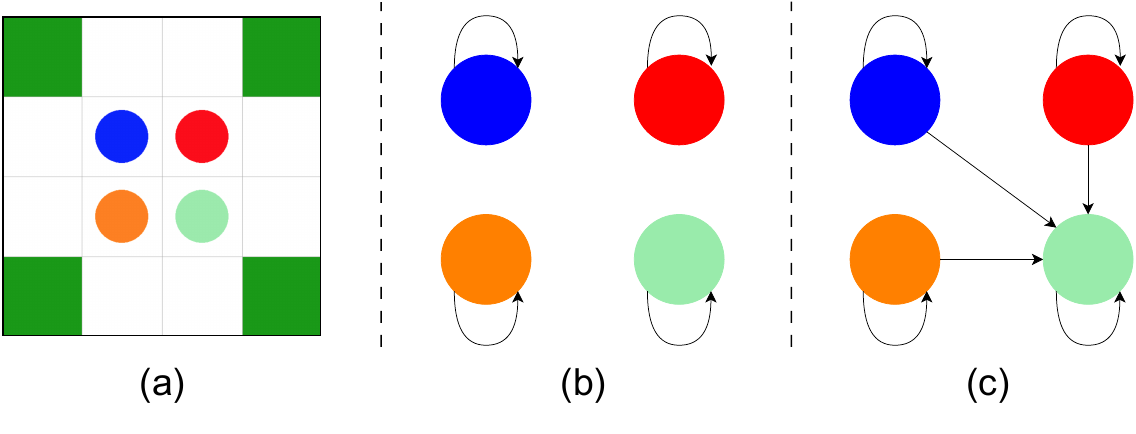}
  \caption{\small{(a) multi-agent grid-world environment with four agents. (b-c) Relational networks employed in CA-VDN.}} 
  \label{fig:discrete-env_and_relations}
\end{figure}

In this environment, depicted in Fig~\ref{fig:discrete-env_and_relations}(a), the goal of each episode is for agents to consume all resources by visiting their locations. Agents can perform five actions: move up, down, left, right, or stay idle. Moreover, they have access to a special action called \textit{push}, which allows them to push adjacent agents. This action is possible when the pushing agent executes a non-idle action towards an idle agent to be pushed. Following a \textit{push}, the agent initiating the push stays stationary, whereas the pushed agent is moved one space in the direction of the push. When an agent successfully consumes a resource, it receives a reward of $+10$. Each resource is single-use and can be consumed by only one agent. However, the agent incurs an individual penalty of $-1$ for each time-step per unconsumed resource, except when occupying a resource location, which acts as a safe spot. The episode ends when either all resources have been consumed or the maximum number of time-steps has been achieved.

The design of this environment is intended to be solvable by VDN, while also highlighting the challenges that unforeseen malfunction(s) can introduce, even in simple scenarios. A malfunction is simulated by immobilizing the green agent. This setup illustrates the impact of integrating agent relationships into the learning to address these challenges.

\subsection{Multi-agent MuJoCo}
\label{mamujoco}
\begin{figure}[b]
\centering
\includegraphics[width=\linewidth]{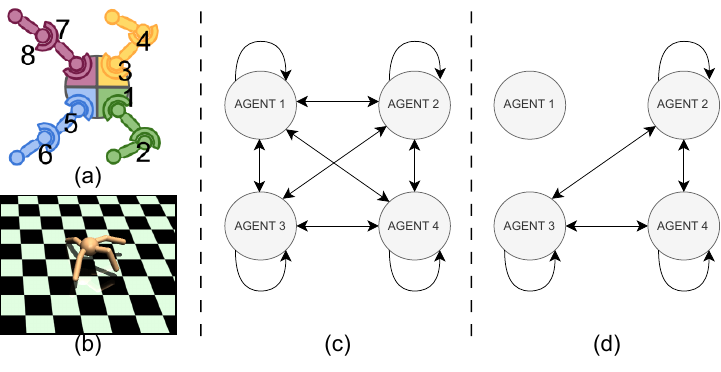}
  \caption{\small{(a) Representation of an ant featuring four agents, each distinguished by a different color and (b) The MaMuJoCo-Ant simulation environment. (c-d) Relational networks used in CA-MQF.}} 
  \label{fig:continuous-env_and_relations}
\end{figure}

\begin{figure*}[b]
\centering
\includegraphics[width=0.8\textwidth]{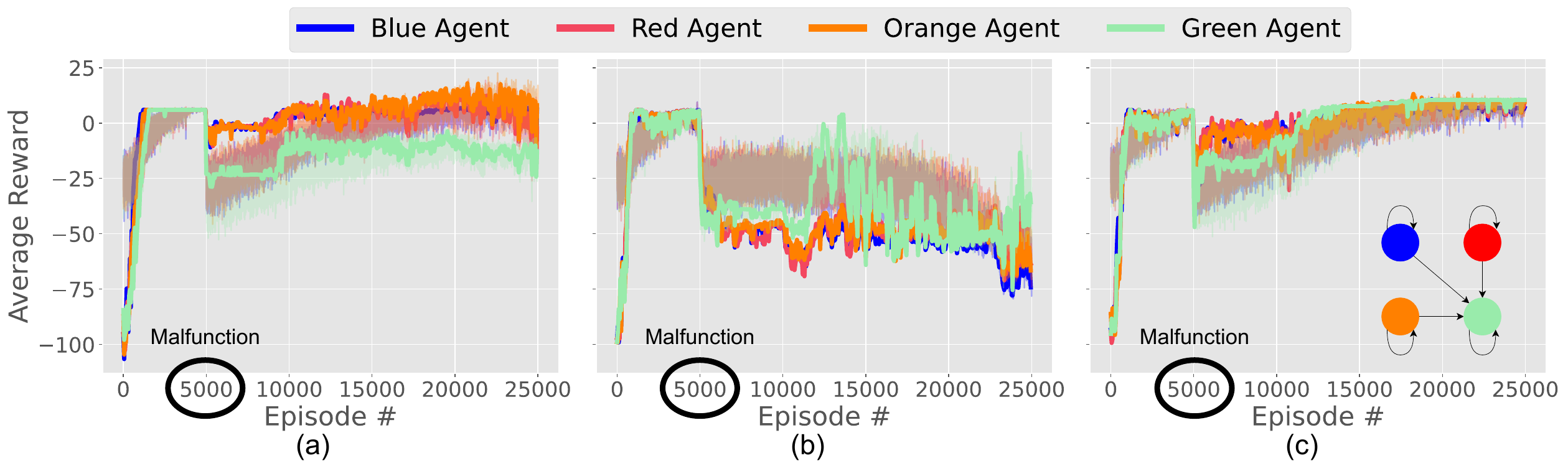}
  \caption{\small{Multi-agent Grid-world results: Average individual rewards of agents before and after the green agent's malfunction at the 5000th episode for (a) IDQN, (b) VDN and (c) CA-VDN.}} 
  \label{fig:discrete-results}
\end{figure*}

Multi-agent MuJoCo (MAMuJoCo)~\cite{peng2021facmac} is a novel benchmark for continuous cooperative multi-agent robotic control. Basically, it is an extension of MuJoCo~\cite{todorov2012mujoco}, by creating a wide variety of novel scenarios in which multiple agents within a single robot have to solve a task cooperatively. More specifically, we used the MaMuJoCo-Ant environment which is a robotic simulation environment that emulates a 3D ant, as depicted in Fig.\ref{fig:continuous-env_and_relations}(b). It consists of a freely rotating torso linked to four legs, with each leg having two joints. The primary goal of this environment is to achieve coordinated movement in the forward direction (toward the positive direction on the $x$-axis). This is achieved by strategically applying torques to the eight joints, thus controlling the ant's movements. 

In our experiments, we employed a variant of the ant environment that assigns an individual agent to each leg, as depicted in Fig\ref{fig:continuous-env_and_relations}(a). Consequently, the ant is controlled by four distinct agents, each in charge of a leg comprising two joints. In this configuration, each agent has two action values that range from $[-1, +1]$. During an episode, agents receive a team reward at each step, calculated as $r_{\textrm{team}} = r_{\textrm{stable}} + r_{\textrm{forward}} - r_{\textrm{ctrl\_cost}}$, where $r_{\textrm{stable}}$ is a fixed reward given at every time-step for maintaining a stable posture (not upside down), $r_{\textrm{forward}}$ is the reward for moving forward, measured as $\frac{\Delta x}{dt}$ (with $dt$ representing the time between actions and $\Delta x$ the change in the $x$ direction), and $r_{\textrm{ctrl\_cost}}$ is the penalty for executing excessively large actions. Additionally, agents are penalized with a $-100$ reward if the ant becomes upside down. For our experiments, we set $r_{\textrm{stable}}$ at $+0.01$ and the maximum number of steps per episode at $100$. It is important to notice that agents have no access to their individual rewards, but only team reward, $r_{\textrm{team}}$.

The integration of multiple agents within a single robotic ant showcases the effectiveness of learning algorithms in enhancing robustness against single-point failures, making it well-suited for our experimental framework. In particular, for our experiments, one leg is immobilized to represent a malfunction. The experiments are designed to evaluate the agents' ability to effectively synchronize actions and adapt to unforeseen malfunction that their teammate(s) might encounter.

\section{Models and Hyperparameters}

In this study, the neural network architecture employed for agent modelling in all algorithms tested is the Multi-Layer Perceptron (MLP) for both environments. For the discrete environment, the MLP configuration consists of two hidden layers, each with 128 neurons, using the ReLU activation function.
The prediction networks are trained at the end of each episode through 10 iterations, utilizing batches of 32 instances (time-steps) that are randomly selected from the replay memory, capable of storing up to 50k time-steps. Every 200 episodes, the weights of the target neural networks are re-assigned with those of prediction networks. In the continuous environment, the MLPs feature three hidden layers, each with 256 neurons, and employ the TanH activation function (while critic networks in MADDPG still uses the ReLU). The prediction networks' weight are updated every 10 steps with batches of 512, randomly chosen from a replay memory of 500k time-steps. Target networks receive updates at each time-step through a \textit{soft} update process, integrating a small factor ($\tau=0.01$) with the weights from the prediction networks. In both discrete and continuous environments, the \textit{Adam} optimizer is utilized for network optimization, and the squared Temporal Difference (TD) error is used as the metric for loss evaluation.

\section{Experimental Results}
In this section, we present our experimental results on MaMuJoCo-Ant and our cooperative multi-agent grid-world environment. We have simulated a malfunction after agents converged to one of the several possible solutions (potentially the global optimal solution). Then, we compared the adaptation abilities of our collaborative adaptation framework with IDQN and VDN for discrete tasks, and with IQF and MADDPG for continuous action domains.

In case of a unforeseen failure, our study assumes that a detection mechanism is in place, capable of determining the malfunction's timing (possibly by monitoring the team's collective reward) and identifying the specific agent that is malfunctioning. The identification process might involve analyzing the agents' previous movements to identify deviations from the behavioral patterns established during training.

\begin{figure*}[b]
\centering
\includegraphics[width=\linewidth]{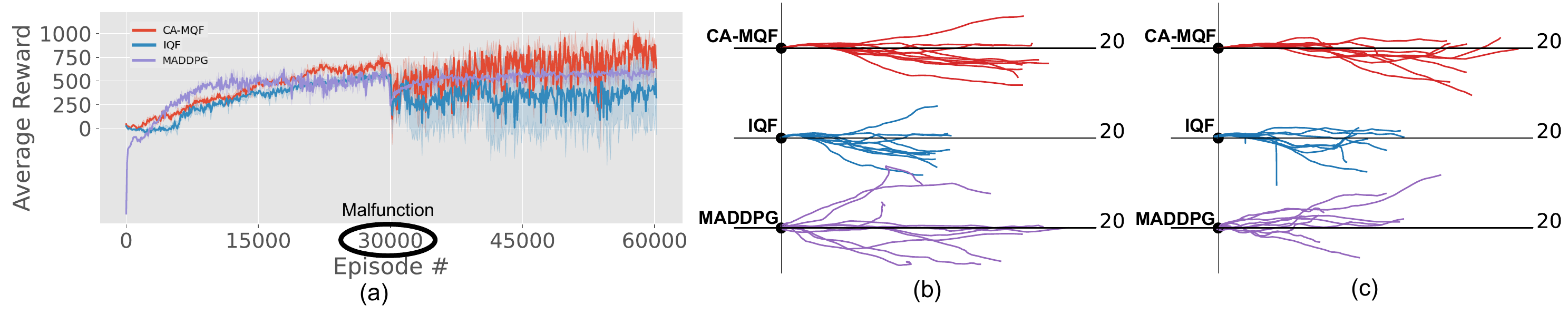}
  \caption{\small{MaMuJoCo-Ant results: (a) Average team rewards before and after malfunction occurred at the $30000$th episode. (b-c) Robot trajectories in x-y plane: (b) before and (c) after malfunction, upon completing 30k and 60k training episodes, respectively. It can be seen that the robot can cover more distance using CA-MQF with higher rewards.}} 
  \label{fig:continuous-results}
\end{figure*}

\vspace{.2cm}
\noindent\textbf{The CA framework demonstrates superior adaptability to unexpected agent failure compared to IDQN and VDN in our multi-agent grid-world environment involving discrete actions.} Despite the initial similarity in performance between the algorithms, IDQN and VDN encounter challenges in recovering post-malfunction, where the green agent, as depicted in Figure 1(a), is immobilized. To improve the algorithms' ability to identify new strategies or behaviors, the epsilon ($\varepsilon$) value is reset after the malfunction incident. Fig.~\ref{fig:discrete-results} displays the experimental outcomes in this setup, showcasing the average training (indicated by shaded areas) and test rewards (represented by solid lines) across ten runs. The test rewards are calculated by assessing the agents' individual rewards using a greedy approach, with the training process being paused after every 50 episodes.

IDQN-trained agents, after malfunction, operate autonomously without a system to foster mutual cooperation, leading to inconsistent behaviors across multiple runs. Following the consumption of three resources, agents typically engage in one of three behaviors: (i) the episode continues due to an unconsumed resource, resulting in random actions and penalties for the agents; (ii) an agent may learn to consume the remaining resource, thereby concluding the episode; (iii) an agent might discover how to push the malfunctioning agent, aiming to secure the reward for pushing while another goes to consume the remaining resource to end the episode. However, there is no evidence of intentional collaborative efforts among the agents, such as assisting the malfunctioning agent in consuming the nearest resource by pushing.

In VDN, despite increased exploration following this malfunction, challenges in recovery persist in some runs, leading to unstable agents (refer to Fig.~\ref{fig:discrete-results}(b)). Specifically, in 40\% of the runs, agents learn to push malfunctioning agents toward a resource and then proceed to the nearest resource. However, in 60\% of the runs, agents either become stuck pushing each other or consume resources without assisting the malfunctioning agent.

On the other hand, the CA framework, employing VDN as its learning algorithm (denoted as CA-VDN), incorporates a relational network as illustrated in Figure~\ref{fig:discrete-env_and_relations}(c). This network enables all agents to prioritize the green agent, identified as the malfunctioning agent, thereby expediting the adaptation process. In this setting, the relational network is directly integrated into the rewards, given that the agents have access to their individual rewards. The observed collaborative behavior, aimed at optimizing team performance, is characterized by the following movements: (i) the red agent pushes the green agent downward by one cell, (ii) the orange agent moves the green agent one cell right. This cooperation allows the green agent to access the nearest resource location despite its malfunction. The framework's effectiveness is validated by a comparative analysis of the individual and total rewards of each agent before and after the malfunction, with the results presented in Table~\ref{discrete_num_results}.

\begin{table}[t]
\centering
\caption{
\small{Average metrics with 95\% confidence intervals for ten runs upon training completion.
}}

\resizebox{\linewidth}{!}{
\begin{tblr}{
  row{even} = {c},
  row{1} = {c},
  row{3} = {c},
  row{5} = {c},
  cell{1}{2} = {c=3}{},
  cell{1}{5} = {c=3}{},
  cell{7}{1} = {c},
  vline{4-5} = {1}{},
  vline{5} = {2-7}{},
  hline{1,7-8} = {-}{},
}
             & Before Malfunction &           &           & After Malfunction &              &                     \\
             & IDQN                & VDN       & CA-VDN    & IDQN               & VDN          & CA-VDN              \\
Blue Agent   &    6.00±0.00                & 5.80±0.25 & 5.50±0.64 &      6.00±2.37              & -74.20±18.14 & \textbf{6.90±0.19}  \\
Red Agent    &             6.00±0.00       & 5.50±0.50 & 5.70±0.40 &     -2.40±12.36               & -63.60±20.26 & \textbf{9.90±0.19}  \\
Orange Agent & 6.00±0.00                   & 5.20±0.67 & 5.40±0.79 &   8.10±7.39                 & -66.50±21.47 & \textbf{9.70±0.40}  \\
Green Agent  & 6.00±0.00                   & 5.70±0.40 & 5.60±0.74 &    -12.70±6.45               & -35.70±28.60 & \textbf{10.50±0.93} \\
Sum          &   24.00±0.00                 &  22.20±1.64         &  22.20±2.38         & -1.00±22.59                  &     -240.00±72.87          &     \textbf{37.00±1.49}                 
\end{tblr}
}

\label{discrete_num_results}
\end{table}

\vspace{.2cm}
\noindent\textbf{The CA framework outperforms both IQF and MADDPG in the MaMuJoCo-Ant environment for continuous action domains.} 
Once the ant is fully capable of navigating the map (up to the 30000th episode), all tested algorithms enable four agents (representing each leg) to collaboratively learn to move the ant in the $+x$ direction. It is crucial to note that, although the environment \textit{positively} rewards such movement, it concurrently imposes penalties for the execution of large actions (refer to $r_\textrm{ctrl\_cost}$  in~\ref{mamujoco}), potentially limiting energy consumption per action. 

In this setting, the CA framework employs QF as a learning algorithm, referred to as CA-MQF. Due to the agents' lack of access to individual rewards, relational networks are applied to their state-action values. As demonstrated in Table~\ref{continuous_num_results}, both CA-MQF, with the relational network depicted in Fig.~\ref{fig:continuous-env_and_relations}(b), and MADDPG exhibit comparable performance, and they surpass IQF in achieving distance from the origin in the $+x$ direction. However, CA-MQF outperforms both MADDPG and IQF in terms of rewards, as MADDPG tends to generate excessively large actions.

In the 30000th episode, a simulated malfunction restricts the movement of one leg, leading to a significant decrease in the average rewards of the algorithms, as depicted in Fig.~\ref{fig:continuous-results}(a). This incident necessitates modification of the relational network from Fig.~\ref{fig:continuous-env_and_relations}(b) to Fig.~\ref{fig:continuous-env_and_relations}(c). The importance placed by non-malfunctioning agents to the malfunctioning one drops to zero (refer to Fig.~\ref{fig:continuous-env_and_relations}(c)), as its state-action values becomes unreliable because of the malfunction. The trajectories in Fig.~\ref{fig:continuous-results}(b-c) illustrate that CA-MQF, with its adjusted inter-agent relationships, is effective in recovering from such malfunctions. While MADDPG outperforms IQF, it does not reach the level of adaptability demonstrated by CA-MQF. The accompanying video shows the animation of this experiment.\footnote{Accompanying video: \url{https://youtu.be/-0Qd5jyRGIY}}

Upon deeper analysis, a noticeable difference is observed between the average team reward depicted in Figure~\ref{fig:continuous-results}(a) and Table~\ref{continuous_num_results}. This difference arises because, to reduce the training time, the testing average team reward was computed with only 100 episodes, pausing training every 1000 steps, whereas the data in Table~\ref{continuous_num_results} were obtained from 10000 test episodes upon training completion, leading to more reliable results. This may also indicate that the IQF and MADDPG models are not stable, as their rewards decreased when the number of test episodes increased.

Overall, the results from experiments in both the discrete and continuous environments highlight that agents trained using the Collaborative Framework exhibit enhanced cooperative behaviors. Furthermore, these agents display an improved capacity for adapting their policies to recover from unexpected malfunctions, capitalizing on the dynamics of inter-agent relationships.

\section{Conclusion and Future Work}
This study presents a novel framework that integrates inter-agent relationships into the learning process, specifically addressing unexpected malfunction scenarios. The efficacy of our method in improving cooperative behaviors among agents and facilitating effective adaptation to unforeseen robotic malfunctions has been demonstrated in both continuous and discrete environments. The experimental findings indicate that the tested algorithms often stuck at sub-optimal solutions, thus decreasing team rewards, whereas our proposed algorithm successfully mitigates and recovers from these malfunctions. Future research will aim to expand the experiments to more complex situations involving multiple agents encountering various unexpected malfunctions (for instance, noisy or adversarial leg/agent) and undertake a comparative analysis with other leading methodologies to further evaluate the robustness and efficiency of our framework in these settings.

\section{Acknowledgments}

This work is supported in part by NSF (IIS-2112633) and
the Army Research Lab (W911NF20-2-0089).

\begin{table}[t]
\centering
\caption{
\small{Average metrics with 95\% confidence intervals for three runs upon training completion.
}}
\resizebox{\linewidth}{!}{
\begin{tblr}{
  cells = {c},
  cell{1}{2} = {c=2}{},
  cell{1}{4} = {c=2}{},
  vline{3-4} = {1}{},
  vline{4} = {2-5}{},
  hline{1,6} = {-}{},
}
                    & Before Malfunction &                 & After Malfunction       &                       \\
       & Team Reward        & Origin Distance & Team Reward             & Origin Distance       \\
IQF    & 568.86 ± 16.06     & 10.37 ± 0.33    & 369.95 ± 204.02         & 8.81 ± 1.65           \\
MADDPG & 296.24 ± 15.64     & 13.81 ± 0.25    & 340.23 ± 20.02          & 11.37 ± 0.40          \\
CA-MQF  & 724.38 ± 19.16     & 13.20 ± 0.43    & \textbf{768.02 ± 14.55} & \textbf{12.80 ± 0.14}      
\end{tblr}
}
\vspace{-5px}
\label{continuous_num_results}
\end{table}

\bibliographystyle{IEEEtran}
\bibliography{IEEEabrv, refs}

\end{document}